\documentstyle[12pt]{article}
\def\CTPa{\it Center for Theoretical Physics, Department of Physics,
      Texas A\&M University}
\def\CTPb{\it College Station, TX 77843-4242, USA}
\def\HARCa{\it Astroparticle Physics Group,
Houston Advanced Research Center (HARC)}
\def\HARCb{\it The Woodlands, TX 77381, USA}

\begin{document}
\vspace{.5cm}
\thispagestyle{empty}
\hfill\vbox{\hbox{\bf CTP-TAMU-08/93}\hbox{\bf PURD-TH-93-1}
\hbox{February 1993}}

\vspace{1.cm}
\begin{center}
{\Large\bf Boson Pair Productions}\\[.25in]
{\Large\bf In $e^+$ $e^-$ Annihilation}\\[1cm]
{\large  T.K.~Kuo$^a$, Gye T.~Park$^{b,c}$ and M. Zralek$^d$}\\[.5cm]
$^a${\it Department of Physics, Purdue University}\\
{\it West Lafayette, IN 47907, USA}\\
$^b${\CTPa}\\
{\CTPb}\\
$^c${\HARCa}\\
{\HARCb}\\
$^d${\it Department of Field Theory and Particle Physics}\\
{\it University of Silesia, PL-40007 Katowice, Poland}\\
\end{center}
\vspace{.75cm}

\begin{abstract} \normalsize
We examine the processes $e^+ e^-\longrightarrow W^+ W^-$ and $Z^0 Z^0$ in the
context of the $SP(6)_L\otimes U(1)_Y$ model. We find that there are
significant deviations in the total cross sections $\sigma (s)$ from the
standard model results due to the presence of additional gauge bosons
$Z^\prime$ and $W^\prime$ in the model. These deviations could be detected at
LEP.
\end{abstract}

\newpage
\setcounter{page}{1}

\section{Introduction}

A major target experiment at the CERN $e^+ e^-$ collider LEPII ($\sqrt s=200$
GeV) is
undoubtedly W-boson pair productions. It should thus be possible to examine
electroweak
(EW) theories quantitatively. The process\cite{eeww}
$e^+ e^-\longrightarrow W^+ W^-$ allows us not
only to determine various properties of the W boson,but to measure
the trilinear couplings\cite{vww}
$VW^+W^-$ ($V=Z^0,\gamma$). Another boson pair production process
$e^+ e^-\longrightarrow Z^0 Z^0$ is physically less important than the
W-pair and $Z\gamma$ pair production because of smaller cross section
but it has to be studied in order to provide a crucial background for
high mass Higgs searches\cite{higgs}. It is still an important
test for the standard model (SM) to examine trilinear couplings.

The aim of this paper is to investigate the possibility that there could be
significant changes in the behavior of the process
$e^+ e^-\longrightarrow W^+ W^-$ and $Z^0 Z^0$ in terms of the total
cross sections $\sigma (s)$ for $\sqrt s\sim 200$ GeV, leading to
considerable deviations from the SM result if a new neutral
gauge boson ($Z^{\prime 0}$) and a new charged gauge boson ($W^{\prime \pm}$)
such as the ones in the
$SP(6)_L\otimes U(1)_Y$ model were present.
There are a number of authors who have investigated this problem within models
with only one extra neutral gauge boson\cite{extraz}.
In fact, two of us (T. K and G. P) presented in a
previous paper a similar work on $e^+ e^-\longrightarrow W^+ W^-$
where the contribution from $W^{\prime \pm}$ was
neglected for simplicity in the analysis. Inclusion of the $W^{\prime \pm}$
turns out to enhance the deviation considerably, as we will show in this paper.

The standard model (SM) has been spectacularly successful in describing
the data that are available from recent experiments\cite{1inUV}. The agreements
between theory and experiments include not just tree-level results, but also
radiative corrections. Nevertheless, there are still a few places where room
for new physics exists. Further, from the theoretical point of view, there is
a consensus that the SM can only be the low energy limit of a more complete
theory. Extensions of the SM usually add extra gauge bosons, or extra fermions,
or both, to the known particle spectrum. In this paper we consider the
$SP(6)_L\otimes U(1)_Y$ family model, in which there is a larger flavor
gauge group with additional gauge bosons, keeping the fermion spectrum
intact.


The $SP(6)_L\otimes U(1)_Y$ model, proposed some time ago\cite{kuo-nakagawa},
is the simplest extension of the standard model of three generations
that unifies the standard $SU(2)_L$ with the horizontal gauge group
$G_H(=SU(3)_H)$ into an anomaly free, simple, Lie group. In this model,
the six left-handed quarks (or leptons) belong to a {\bf 6} of
$SP(6)_L$, while the right-handed fermions are all singlets. It is thus
a straightforward generalization of $SU(2)_L$ into $SP(6)_L$, with the three
doublets of $SU(2)_L$ coalescing into a sextet of $SP(6)_L$. Most of the
new gauge bosons are arranged to be heavy $(\geq 10^2$--$10^3\rm\,TeV)$ so as
to avoid sizable FCNC. $SP(6)_L$ can be naturally broken into $SU(2)_L$
through a chain of symmetry breakings. The breakdown
$SP(6)_L \rightarrow [SU(2)]^3 \rightarrow SU(2)_L$ can be induced by two
antisymmetric Higgs which transform as $({\bf 1}, {\bf 14}, 0)$ under
$SU(3)_C\otimes SP(6)_L\otimes U(1)_Y$. The standard $SU(2)_L$ is to be
identified with the diagonal $SU(2)$ subgroup of
$[SU(2)]^3=SU(2)_1\otimes SU(2)_2\otimes SU(2)_3$, where $SU(2)_i$ operates
on the $i$th generation exclusively. In terms of the $SU(2)_i$ gauge boson
$\vec{A}_i$, the $SU(2)_L$ gauge bosons are given by $\vec{A}={1\over\sqrt 3}
(\vec{A}_1+\vec{A}_2+\vec{A}_3)$. Of the other orthogonal combinations of
$\vec{A}_i$,
$\vec{A}^\prime={1\over\sqrt 6}(\vec{A}_1+\vec{A}_2-2\vec{A}_3)$, which
exhibits unversality only
among the first two generations, can have a mass scale in the TeV range
\cite{1TeVZ}. The three gauge bosons $A^\prime$ will be denoted as $Z^\prime$
and $W^{\prime\pm}$.

\section{The $SP(6)_L\otimes U(1)_Y$ family model and the cross sections}

We now turn to a detailed analysis of the effects of the extra bosons from
$SP(6)_L\otimes U(1)_Y$ model. The dominant effects of new heavier gauge boson
$Z^\prime (W^{\prime\pm})$ show up
in its mixing with the standard $Z(W^\pm)$ to form the mass eigenstates
$Z_{1,2} (W_{1,2})$:
\[ \hbox to \hsize{$ \hfill
\begin{array}{rcl}
Z_1&=&Z\cos\phi_Z+Z^\prime\sin\phi_Z \;, \\
W_1&=&W\cos\phi_W+W^\prime\sin\phi_W \;,
\end{array} \quad
\begin{array}{rcl}
Z_2 &=& -Z\sin\phi_Z+Z^\prime\cos\phi_Z \;, \\
W_2 &=& -W\sin\phi_W+W^\prime\cos\phi_W \;,
\end{array} \hfill
\begin{array}{r}
\stepcounter{equation}(\theequation)\\
\stepcounter{equation}(\theequation)
\end{array}
$} \]
where $Z_1 (W_1)$ is identified with the physical $Z(W)$.

With the additional gauge boson $Z^\prime$, the neutral-current Lagrangian
is generalized to contain an additional term
\begin{equation}
L_{NC}=g_Z J_Z^\mu Z_\mu +g_{Z^\prime} J_{Z^\prime}^\mu Z_\mu^\prime \;,
\end{equation}
where $g_{Z^\prime}=\sqrt{1-x_W\over 2} g_Z={g\over\sqrt{2}}$,
$x_W=\sin^2\theta _W$, and $g={e\over {\sin\theta _W}}$. The neutral currents
$J_Z$ and $J_{Z^\prime}$ are given by
\begin{eqnarray}
J_Z^\mu &=&\sum_{f} \bar{\psi}_f\gamma^\mu\left( g^f_V+g^f_A\gamma _5\right)
\psi_f \;, \\
J_{Z^\prime}^\mu &=&\sum_{f} \bar{\psi}_f\gamma^\mu\left( g^{\prime
f}_V+g^{\prime f}_A\gamma _5\right)
\psi_f \;,
\end{eqnarray}
where $g^f_V={1\over 2}\left( I_{3L}-2x_Wq\right)_f$, $g^f_A={1\over
2}\left( I_{3L}\right)_f$ as in SM, $g^{\prime f}_V=g^{\prime
f}_A={1\over 2}\left( I_{3L}\right)_f$
for the first two generations and $g^{\prime f}_V=g^{\prime
f}_A=-\left( I_{3L}\right)_f$ for the third. Here $\left(I_{3L}\right)_f$ and
$q_f$
are the third component of weak isospin and electric charge of fermion $f$,
respectively. And the neutral-current Lagrangian reads in terms of $Z_{1,2}$
\begin{equation}
L_{NC}=g_Z\sum_{i=1}^2\sum_{f} \bar{\psi}_f\gamma_\mu\left(
g^f_{Vi}+g^f_{Ai}\gamma _5\right)
\psi_f Z^\mu_i \;,
\end{equation}
where $g^f_{Vi}$ and $g^f_{Ai}$ are the vector and axial-vector
couplings of fermion $f$ to physical gauge boson $Z_i$, respectively.
They are given by
\begin{eqnarray}
g^f_{V1, A1}&=&g^f_{V, A}\cos\phi_Z+{g_{Z^\prime}\over g_Z} g^{\prime
f}_{V, A}\sin\phi_Z \;, \\
g^f_{V2, A2}&=&-g^f_{V, A}\sin\phi_Z+{g_{Z^\prime}\over g_Z} g^{\prime
f}_{V, A}\cos\phi_Z \;.
\end{eqnarray}
Similar analysis can be carried out in the charged sector.

In order to see visible effects of the presence of $Z^{\prime}$ and
$W^{\prime}$,
the mixing angles $\phi_Z$ and $\phi_W$ should not be too small.
According to an analysis\cite{tests} using the latest
LEP data the present constraint on the mixing angles is
$\vert\phi_Z\vert$, $\vert\phi_Z\vert\leq 0.01$ .
This constraint will be used in choosing the mixing angles in the following
analysis.

Let's first consider the process $e^+ e^-\longrightarrow W_1^+ W_1^-$
$$e^+(p_+,\sigma_+) + e^-(p_-,\sigma_-)\longrightarrow
W_1^+(k_+,\lambda_+) + W_1^-(k_-,\lambda_-)
$$
Neglecting the electron mass for high energies, only
two initial helicity configurations $\Delta\sigma =\sigma_- -\sigma_+ =\pm 1$
are allowed. All polarized differential cross sections are given by
%
\begin{equation}
{d\sigma_{\Delta\sigma,  \lambda_+ \lambda_-}\over d \cos\theta}=
{x\over 16\pi s} \vert M(\Delta\sigma,  \lambda_+ \lambda_-)\vert^2
\end{equation}
where $\theta$ is the angle between the $e^-$ and the $W_1^-$ momenta,
$k_- = (E_-,k \sin\theta,0,k \cos\theta)$, $x={k\over\sqrt s}$ and
\begin{eqnarray}
M(\Delta\sigma,  \lambda_+ \lambda_-)&=&- {e^2\over\sqrt 2}
{d^{J_0}_{\Delta\sigma,  \Delta\lambda}}(\theta)
\left[
{2 X^{\Delta\sigma}_t\over {A+4 x \cos\theta}} M_t
(\Delta\sigma,  \lambda_+ \lambda_-)- M_s (\Delta\sigma,  \lambda_+ \lambda_-)
\right.\\
\nonumber
&~& \left. \left\{
 X^{\Delta\sigma}_{\gamma}+\right. 
  \hspace*{.20in} \left. X^{\Delta\sigma}_{Z_1} {s\over
 {s-M^2_{Z_1}+i M_{Z_1} \Gamma_{Z_1}}}+
X^{\Delta\sigma}_{Z_2} {s\over
 {s-M^2_{Z_2}+i M_{Z_2} \Gamma_{Z_2}}}
 \right\}
\right]
\end{eqnarray}
%
where $\Delta\lambda=\lambda_- - \lambda_+$,
$J_0= max(\vert\Delta\sigma\vert, \vert\Delta\lambda\vert)$,
${d^{J_0}_{\Delta\sigma,  \Delta\lambda}}(\theta)$ being an ordinary Wigner
function, and
\begin{eqnarray}
A&=& -(1+4x^2), \quad X^{\Delta\sigma}_{\gamma}=-1              \;,\\
X^{\Delta\sigma =-1}_t&=&2 B_L^2, \quad X^{\Delta\sigma =+1}_t= 0 \;,\\
X^{\Delta\sigma=-1}_{Z_1} &=& A^1_{L_e} \Lambda_{Z_1},
\quad X^{\Delta\sigma=+1}_{Z_1} = A^1_{R_e} \Lambda_{Z_1} \;,\\
X^{\Delta\sigma=-1}_{Z_2} &=& A^2_{L_e} \Lambda_{Z_2},
\quad X^{\Delta\sigma=+1}_{Z_2} = A^2_{R_e} \Lambda_{Z_2} \;.
\end{eqnarray}
where

\begin{eqnarray}
B_L&=& {1\over{\sqrt 2} s} (\cos\phi_W+ {1\over\sqrt 2}\sin\phi_W) \;,\\
A^1_{L_e}&=& {1\over 2 s c}\left[ (-1+2 s^2)\cos\phi_Z-
{c\over\sqrt 2}\sin\phi_Z\right] \;,\\
A^2_{L_e}&=& {1\over 2 s c}\left[ (1-2 s^2)\sin\phi_Z-
{c\over\sqrt 2}\cos\phi_Z\right] \;,\\
A^1_{R_e}&=& ({s\over c})\cos\phi_Z \;, A^2_{R_e}= -({s\over c})\sin\phi_Z
\;,\\
\Lambda_{Z_1}&=& {1\over s}(c\cos\phi_Z-
{1\over\sqrt 2}\sin\phi_Z\sin^2\phi_W) \;,\\
\Lambda_{Z_2}&=& -{1\over s}(c\sin\phi_Z+
{1\over\sqrt 2}\cos\phi_Z\sin^2\phi_W) \;,
\end{eqnarray}

with $s\equiv\sin\theta_W$, $c\equiv\cos\theta_W$.
Similarly, for the process $e^+ e^-\longrightarrow Z^0_1 Z^0_1$,
\begin{equation}
M(\Delta\sigma,  \lambda_+ \lambda_-)=- 2\sqrt 2e^2
{d^{J_0}_{\Delta\sigma,  \Delta\lambda}}(\theta) X^{\Delta\sigma}\left[
{M_t (\Delta\sigma,  \lambda_+ \lambda_-)\over {A+4 x \cos\theta}} -
{M_u (\Delta\sigma,  \lambda_+ \lambda_-)\over {A-4 x \cos\theta}}\right]
\end{equation}
where $X^{\Delta\sigma=-1}=(A^1_{L_e})^2$ and
$X^{\Delta\sigma=+1}=(A^1_{R_e})^2$. For brevity we omit the explicit
expressions for $M_t$,$M_u$,and $M_s$ which are t-, u- and s- channel
amplitudes, respectively. Using the above formulas and
$\sin^2\theta_W = 0.23$ , $M_{Z_1}=91.17$ GeV, $M_{W_1}=80.11$ GeV,
$\Gamma_{Z_1}=2.5$ GeV and the calculated value for
$\Gamma_{Z_2}$\cite{eewwsp6}, we calculate the total cross sections for
$e^+ e^-\longrightarrow W_1^+ W_1^- , Z^0_1 Z^0_1$.

Now let us turn to our numerical results.
Figures 1 and 2 show $\sigma(e^+ e^-\longrightarrow W_1^+ W_1^-)$ for four
different sets of mixing angles and a fixed $M_{Z_2}$ in comparison with the SM
results. We see that the effects of
the extra gauge bosons is more pronounced for $\phi_Z=-\phi_W$. The deviations
of $\sigma$ from the SM result at $\sqrt s =200$ GeV are $2.94-3.34\%$
for $\vert\phi_Z\vert = \vert\phi_W\vert = 0.01$ and $M_{Z_2}=1$ TeV.
Considering the fact that the deviations were found to be less than $1.1\%$
for $\vert\phi_Z\vert = 0.05$ and $M_{Z_2}=500$ GeV neglecting $W^\prime$
contribution\cite{eewwsp6}, it is very interesting to see that there is
considerable contribution from the charged sector.
Therefore, an accurate measurement (with statistical error $\leq 1\%$) of
$\sigma$ at $\sqrt s =200$ GeV at LEP II will be able to test the
$SP(6)_L\otimes U(1)_Y$ model at the level of $\vert\phi_Z\vert\simeq
\vert\phi_W\vert\simeq 0.01$.
Figure 3 shows $\sigma(e^+ e^-\longrightarrow Z^0_1 Z^0_1)$ for $\phi_Z=\pm
0.01$ and $M_{Z_2}=1$ TeV. The deviations from the SM result at $\sqrt s =200$
GeV are $\sim 3.0\%$.

\section{Summary and Conclusions}

We have examined the processes $e^+ e^-\longrightarrow W^+ W^-$ and $Z^0 Z^0$
in the context of the $SP(6)_L\otimes U(1)_Y$ model. Owing to the presence of
the additional gauge bosons $Z^\prime$ and $W^\prime$ in this model,
total cross section $\sigma (s)$ can be significantly different from that of
the standard model. This effect is dependent on the mixing angles between
$Z(W)$ and $Z^\prime(W^\prime)$. For mixing angles at the level of $1\%$,
these deviations are roughly $3\%$, which should be detectable at LEPII.
Thus, the production of $W$ and $Z$ pairs should provide a sensitive test of
possible new physics beyond the SM.

\section*{Acknowledgements}

T.~K. was supported in part by the U.S.~Department of Energy under Contract
No.~DE-AC02-76ER01428. G.~P. was supported by an ICSC-World Laboratory
Scholarship.

\newpage
\section*{Figure Captions}

\begin{enumerate}

\item The total cross section $\sigma(e^+ e^-\longrightarrow W_1^+ W_1^-)$
as a function of $\sqrt s$ for $SP(6)_L\otimes U(1)_Y$ model in comparison with
the SM value for $M_{Z_2}=1$ TeV. Solid line: SM; long dashed:
$\phi_Z=\phi_W=0.01$;
short dashed: $\phi_Z=\phi_W=-0.01$.

\item Same as in Figure 2 except for $\phi_Z=-\phi_W=\pm 0.01$ used instead.

\item The total cross section $\sigma(e^+ e^-\longrightarrow Z^0_1 Z^0_1)$
as a function of $\sqrt s$ for $SP(6)_L\otimes U(1)_Y$ model in comparison with
the SM value for $M_{Z_2}=1$ TeV. Solid line: SM; long dashed: $\phi_Z=0.01$;
short dashed: $\phi_Z=-0.01$.

\end{enumerate}

\end{document}